\documentclass[aps,prl,twocolumn,superscriptaddress]{revtex4-2}

\usepackage{graphicx}
\usepackage{hyperref}
\usepackage{amsmath}
\usepackage{amsfonts}
\usepackage{ulem}

\hypersetup{
	colorlinks=true,
	linkcolor=blue,
	citecolor=blue,
	urlcolor=blue
}

\begin{document}
	
\title{Observation of strong and tunable light-induced dipole-dipole interactions between optically levitated nanoparticles}	
	
	\author{Jakob Rieser}
	\author{Mario A. Ciampini}
	\affiliation{University of Vienna, Faculty of Physics, Vienna Center for Quantum Science and Technology (VCQ), Boltzmanngasse 5, A-1090 Vienna, Austria}
	\author{Henning Rudolph}
	\affiliation{University of Duisburg-Essen, Faculty of Physics, Lotharstraße 1, 47048 Duisburg, Germany}
	\author{Nikolai Kiesel}
	\affiliation{University of Vienna, Faculty of Physics, Vienna Center for Quantum Science and Technology (VCQ), Boltzmanngasse 5, A-1090 Vienna, Austria}
	\author{Klaus Hornberger}
	\author{Benjamin A. Stickler}
	\email{benjamin.stickler@uni-due.de}
	\affiliation{University of Duisburg-Essen, Faculty of Physics, Lotharstraße 1, 47048 Duisburg, Germany}
	\author{Markus Aspelmeyer}
	\affiliation{University of Vienna, Faculty of Physics, Vienna Center for Quantum Science and Technology (VCQ), Boltzmanngasse 5, A-1090 Vienna, Austria}
	\affiliation{Institute for Quantum Optics and Quantum Information (IQOQI) Vienna, Austrian Academy of Sciences, Boltzmanngasse 3, A-1090 Vienna, Austria}
	\author{Uro\v s Deli\' c}
	\email{uros.delic@univie.ac.at}
	\affiliation{University of Vienna, Faculty of Physics, Vienna Center for Quantum Science and Technology (VCQ), Boltzmanngasse 5, A-1090 Vienna, Austria}

\date{\today}

\begin{abstract}
	
Arrays of optically trapped nanoparticles have emerged as a promising platform for the study of complex non-equilibrium phenomena. Analogous to atomic many-body systems, one of the crucial ingredients is the ability to precisely control the interactions between particles. However, the optical interactions studied thus far only provide conservative optical binding forces of limited tunability. Here we demonstrate a coupling mechanism that is orders of magnitude stronger and has new qualitative features. These effects arise from the previously unexplored phase coherence between the optical fields that drive the light-induced dipole-dipole interaction. In addition, polarization control allows us to observe electrostatic coupling between charged particles in the array. Our results pave the way for a fully programmable many-body system of interacting nanoparticles with tunable dissipative and nonreciprocal interactions, which are instrumental for exploring entanglement and topological phases in arrays of levitated nanoparticles.
	
\end{abstract}


\maketitle

When a dielectric sub-wavelength particle is illuminated by laser light, it polarizes the particle in phase with the incoming electromagnetic wave. The induced dipole makes the particle a \textit{high-field seeker}, which enables optical trapping in the intensity maximum of focused lasers \cite{Ashkin1986}. The dipole radiation field accordingly acquires the optical phase of the trapping field. This process, called \textit{coherent scattering}, has been used in combination with an optical cavity to cool the motion of atoms and polarizable nanoparticles in far-detuned traps \cite{Horak1997,Vuletic2000,Vuletic2001,Leibrandt2009,Hosseini2017,DelicPRL,Windey2018}. More recently, it has been exploited in order to achieve quantum-limited detection and motional ground state cooling of single silica nanoparticles in an optical cavity \cite{DelicScience} and with real-time feedback \cite{Magrini,Tebbenjohanns}. 

Simultaneous trapping of more than one particle in a single optical potential allows for the creation of a self-organized structure of particles interacting through the scattered light \cite{ZemanekRMP}. The particles assume steady-state positions at locations where the constructive interference of scattered fields is maximized, such that the total energy is minimized. This optical interaction is fundamentally conservative and reciprocal, giving rise to a spring-type interaction called {\it optical binding}. Optical binding between dielectric objects has been realized for microparticles (radius $\gg$ wavelength$/10$) in many experiments \cite{BurnsBinding,MonikaRitschBinding,TatarkovaBinding,Mohanty2004,WeiLateralBinding,Bykov2018,AritaBinding,ZemanekTractorBeam,Svak2021Optica}. However, the light scattered from one of these particles, as described by Mie scattering theory \cite{ZemanekRMP}, is not spatially coherent over the extension of a neighboring particle. Furthermore, interparticle coupling due to the ambient gas or liquid (\textit{aero/hydrodynamic coupling}) can be dominant for large objects and small distances, which gives rise to complex particle dynamics in which it is hard to isolate the optical interaction. For the case of nanoscale objects optical binding has been explored for metal particles in liquid, where plasmon resonances enhance the interaction \cite{Zhangmetal,KallMetal,Demergis2012}. Although arrays of optically levitated dielectric nanoparticles are a prime system for investigating complex non-equilibrium phenomena \cite{LechnerMultiple,RitschAdaptiveMultifrequency,HolzmannSelfOrdering,RitschBroadband,RitschDrivenDipole,TongcangParticleChain}, controllable optical interactions between such particles remain unexplored. Tools presented in this article pave the way to utilizing the protocols of atomic physics for the generation and observation of quantum correlations and topological phases in an array of nanoscale dielectrics \cite{LukinQuantumSimulator,BrowaeysQuantumSimulator,TopologyMarquardt,TopologyZoller,TopologicalHuber,TopologyCoulais}.

\begin{figure*}
	\centering
	\includegraphics[width=0.8\linewidth]{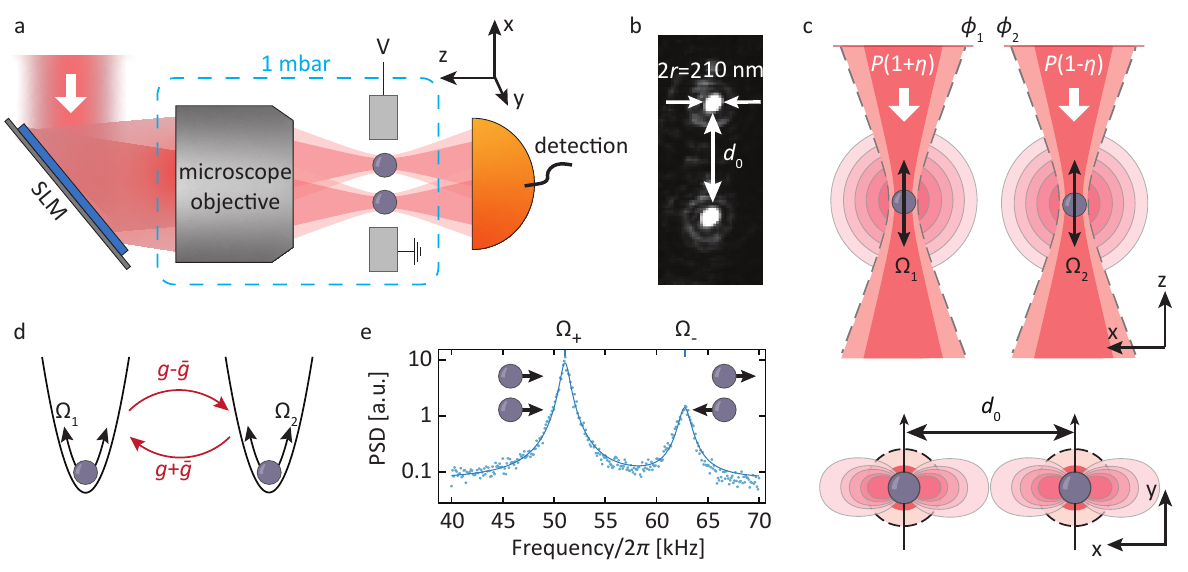}
	\caption{\label{fig:setup} \textbf{Experimental setup.} (a) Two laser beams are diffracted by the spatial light modulator (SLM) and focused with the microscope objective to create two distinct optical tweezers. The optical traps are in the vacuum chamber at a pressure of $\sim 1$ mbar. The light is collected after the focus and used for detection of mechanical modes. (b) Camera image of two nanoparticles (radius $r=105$ nm) trapped in two optical traps at a distance $d_0\sim 10~\mu\text{m}$. (c) Side (above) and top view (below) of the trap foci. Two parallel laser beams are used to trap two nanoparticles at a distance $d_0$. Intrinsic mechanical frequencies along the z axis $\Omega_1\propto \sqrt{P(1+\eta)}$ and $\Omega_2 \propto \sqrt{P(1-\eta)}$ are controlled by a single parameter $\eta$. Polarization is set along the y axis in order to maximize dipole radiation along the x axis. We set the optical phases $\phi_1$ and $\phi_2$ with the SLM. (d) Two harmonic oscillators with frequencies $\Omega_{1,2}$ are coupled via dipole-dipole interactions with non-reciprocal coupling rates $g\pm\bar{g}$. (e) In case of reciprocal couplings ($\bar{g}\equiv 0$) the normal modes of the system -- center-of-mass (CoM) $z_+$ and breathing modes $z_-$ -- are non-degenerate with frequencies $\Omega_+=\Omega$ and $\Omega_-=\Omega+2g$, respectively. We observe the normal modes at frequencies $\Omega_+/2\pi\approx 51$ kHz and $\Omega_-/2\pi\approx 63$ kHz in the power spectral density (PSD) of the detector signal.}
\end{figure*}

In stark contrast to previous experiments, here we levitate two silica nanoparticles with radii significantly smaller than the wavelength (radius $r=105\pm 3$ nm, wavelength $\lambda=1064$ nm) in two distinct, phase-coherent optical traps at a variable trap separation $d_0$. Each particle experiences a total optical field given by the trapping field, effectively inducing a dipole moment, and the coherently scattered light from the other particle. The interference between these two fields gives rise to the interaction between the particles and affects their displacement in all three dimensions. The total light-induced interaction is a combination of a conservative gradient force and a non-conservative radiation pressure force, in analogy to the forces acting on a single nanoparticle in an optical trap \cite{Ashkin1986}. The resulting optical inter-particle forces oscillate periodically and decay as $F_{1,2}\propto \cos (kd \pm \Delta\phi)/kd$ in the far field ($kd \gg 1$, $k=2\pi/\lambda$), where $d$ is the interparticle distance, the wavelength determines the period, and $\Delta\phi$ denotes the optical phase difference between the trapping lasers at the particle positions \cite{SI}. The interaction is fundamentally non-reciprocal ($F_1\neq -F_2$) for $\Delta \phi\neq 0$, contrary to previous experimental optical binding studies where only conservative interactions have been explored ($\Delta \phi\equiv 0$).

We obtain the coupling between the particle displacements along the trap axes ($x,y,z$) from expanding the optical forces in terms of the relative motion. The particle motion modifies the distance $d^2=(d_0+x_1-x_2)^2+(y_2-y_1)^2+(z_2-z_1)^2$. The coupling rate between the x motions thus depends on the distance as $\propto (kd_0)^{-1}$, which has been confirmed in several studies of the lateral optical binding with microparticles \cite{BurnsBinding,Mohanty2004,Demergis2012,AritaBinding,SI}. Using the same argument, one would naively expect that the coupling rate between the z motions would scale with $\propto (kd_0)^{-2}$. However, in our platform the dominant contribution to the coupling stems from the phase dependence of the interference between the local trapping and scattered fields as the motions are encoded in the phase $\Delta\phi = \Delta \phi_0+(k-1/z_R)(z_1-z_2)$ \cite{DelicPRL}, where $\Delta\phi_0$ is the optical phase difference between the trapping lasers in the focal plane and $z_R$ is the Rayleigh length. This results in a long-range coupling rate that depends on the distance as $\propto (kd_0)^{-1}$, thus being a factor of $kd_0\sim 18$ times larger at $d_0=3\lambda$ than the next leading-order contribution. Furthermore, it is effectively larger than the coupling between the x motions by the ratio of mechanical frequencies $\sim 4$, thus making it compelling to explore interaction along the z axis. Altogether, attaining precise control over the optical phases allows us to realize non-reciprocal and ultra-strong dipole-dipole interaction between nanoscale dielectric objects for the first time. 

The linearized dynamics with particle center-of-mass positions $z_{1,2}$ follow as \cite{SI}:
\begin{eqnarray}
	m\ddot{z}_1 +m\gamma\dot{z}_1&=& - \left( m\Omega_1^2 + k_1 + k_2 \right) z_1 + (k_1 + k_2)z_2 \nonumber\\
	m\ddot{z}_2 +m\gamma\dot{z}_2&=& - \left( m\Omega_2^2 + k_1 - k_2 \right) z_2 + (k_1 - k_2) z_1,\label{Langequation}
\end{eqnarray}
where $m$ is the mass of both particles. The spring constant $k_1 = G \cos(kd_0) \cos(\Delta\phi_0)/kd_0$ describes the conservative part of the optical forces (tunable optical binding), while $k_2 = G \sin(kd_0) \sin(\Delta\phi_0)/kd_0$ describes a non-conservative interaction, as indicated by a change of sign between the equations. The constant $G\propto \alpha^2 \sqrt{P_1 P_2}$ is a positive function of the trap powers $P_{1,2}=(1\pm\eta)P$ with the control parameter $\eta$ and the total power $2P$, and the particle polarizability $\alpha$. The scaling of $G$ with $\alpha^2$ reflects the nature of the dipole-dipole interaction. At the pressures in our experiment mechanical damping $\gamma$ is dominated by the collisions with the surrounding gas. For weak coupling between the particles ($k_1\text{, }k_2\ll m\Omega_{1,2}^2$) Eq. \eqref{Langequation} yields the eigenfrequencies of the coupled system $\Omega_\pm\approx\Omega(1-\eta_{\rm m})+g+\bar{g}\mp\sqrt{g^2-\bar{g}^2}$, where we define the conservative and non-conservative coupling rates as $g=k_1/2m\Omega(1-\eta_{\rm m})$ and $\bar{g}=k_2/2m\Omega(1-\eta_{\rm m})$, respectively. The control parameter $\eta_{\rm m}=k_2/2m\Omega^2$ defines the value at which the frequency splitting $\Omega_+-\Omega_-$ is minimal and $\Omega$ is the intrinsic mechanical frequency in absence of interactions at $\eta=0$. The non-reciprocal coupling rates are $g\pm \bar{g}$, thus the distance $d_0$ and the optical phase difference $\Delta\phi_0$ allow for full control over non-reciprocal dipole-dipole interactions between the particles.

In the experiment, the phase-coherent trapping lasers are generated in the first order diffraction of a spatial light modulator (SLM, Meadowlark Optics 512x512 pixels). The lasers are focused by a microscope objective (CFI TU Plan Fluor EPI 50x, Nikon Corp., numerical aperture $\text{NA}=0.8$) into two independent traps (Fig. \ref{fig:setup}a,b). The total trapping power of $2P\sim 800$ mW is split between the two traps as $P(1\pm\eta)$, which allows us to modify the mechanical frequencies along the z axis as $\Omega_{1,2}\propto \sqrt{P(1\pm\eta)}$ (Fig. \ref{fig:setup}c). We control $\eta$, the optical phases $\phi_{1,2}$ at each trapping site and the trap separation $d_0$ (distance between the trap foci along the x axis) with the SLM. As the dipoles do not radiate light along the polarization axis, we set the laser polarization along the y axis in order to maximize the scattering along the x axis and hence the interaction strength. Each particle is randomly charged, therefore we calibrate the absolute charge by applying an AC voltage to two razor-blade electrodes placed along the x axis and select particles based on the desired charge \cite{SI}. We monitor the particle motion with homodyne detection of the light transmitted from the optical traps, which has the particle motion encoded in its phase. In general, our system is described as two harmonic oscillators with frequencies $\Omega_{1,2}$ mutually coupled with non-reciprocal coupling rates $g\pm\bar{g}$ (Fig. \ref{fig:setup}d). In the case of purely conservative interaction ($\bar{g}\equiv 0$) the normal modes of the system become the center-of-mass mode (CoM) $z_+=z_1+z_2$ and the breathing mode $z_-=z_1-z_2$. Only the breathing mode is affected by the interaction, such that its eigenfrequency shifts to $\Omega_-\approx \Omega+2g$, while the CoM mode eigenfrequency remains unchanged at $\Omega_+=\Omega$ (normal mode splitting). In Fig. \ref{fig:setup}e we show the normal modes at $\Omega_+/2\pi\approx 51$ kHz and $\Omega_-/2\pi\approx 63$ kHz in the power spectral density (PSD) of the detector signal, where the mechanical frequency is $\Omega/2\pi\approx 51$ kHz when the interaction is switched off.

\begin{figure}[!htb]
	\centering
	\includegraphics[width=0.75\linewidth]{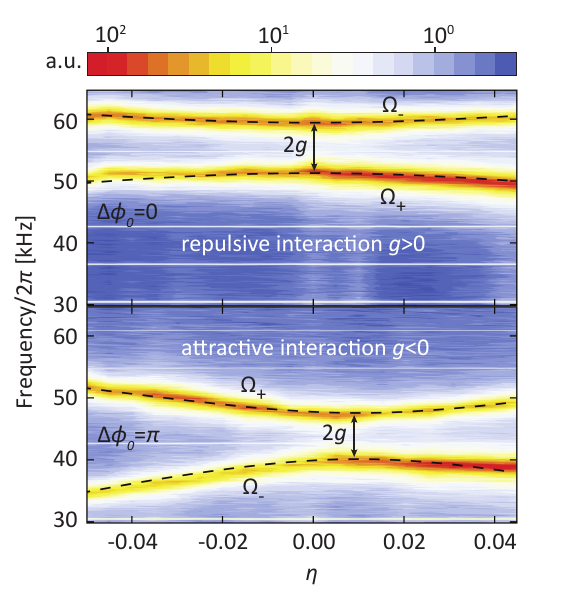}
	\caption{ \textbf{Avoided crossing between the breathing and the CoM mode.} We plot a spectrogram of the relevant frequency region as a function of the control parameter $\eta$. Repulsive and attractive conservative dipole-dipole interactions are observed at the trap separation of $d_0\approx 3.15~\mu\text{m}$ and for the optical phase difference $\Delta\phi_0=0$ (above) and $\Delta\phi_0=\pi$ (below), respectively. The CoM mode is always at the frequency $\Omega_+/2\pi\approx 50$ kHz, while the breathing mode $\Omega_-$ has a higher (lower) frequency by $2g/2\pi\approx 8$ kHz for $\Delta\phi_0=0$ ($\Delta\phi_0=\pi$). Black dashed lines are fits to experimental data.}
	\label{fig:nms}
\end{figure}

\begin{figure}[!htb]
	\centering
	\includegraphics[width=0.75\linewidth]{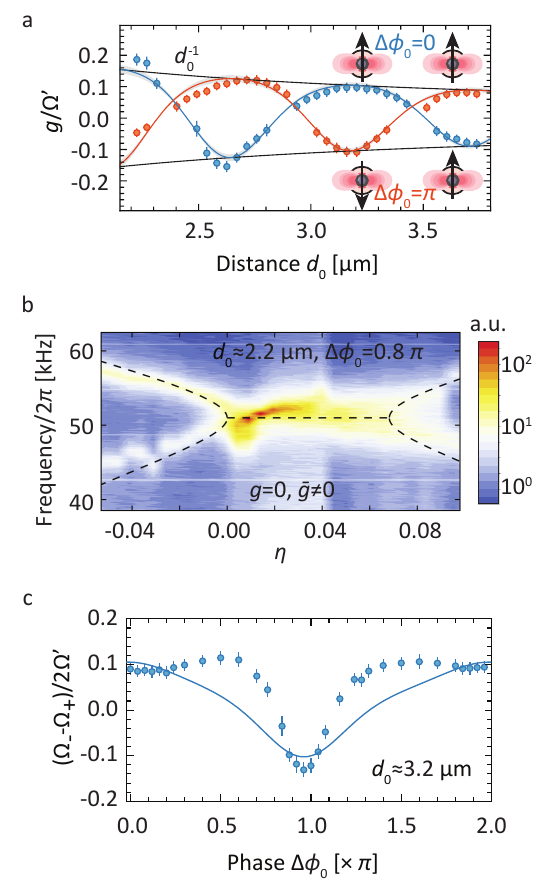}
	\caption{\label{fig:couplingscans} \textbf{Controllable dipole-dipole coupling.} (a) Trap separation $d$ is changed while keeping the optical phase difference fixed at $\Delta\phi_0=0$ (blue circles) or $\Delta\phi_0=\pi$ (orange circles). We observe a change of the coupling rate $g$ with periodicity $\sim \lambda$ and an envelope that drops off as $d_0^{-1}$. Amplitudes of blue and orange lines are calculated from the system parameters with the grayed region given by the standard deviation of the particle size. (b) At the trap separation $d_0\approx 2.2~\mu\text{m}$ the particles experience a combination of the conservative and non-conservative forces. For the optical phase difference of $\Delta\phi_0=0.8\pi$ only the non-conservative interaction is present ($g=0$) and the eigenmodes are degenerate for $\eta$ between $0$ and $0.07$. The motion is strongly amplified in this region. The dashed lines are theory based on the estimated conservative and non-conservative coupling rates. (c) We set the trap separation at $d_0\approx 3.2~\mu\text{m}$ and tune the optical phase difference $\Delta\phi_0$ from $0$ to $2\pi$ and measure the mode splitting $\Omega_--\Omega_+$. Interaction is mostly conservative at $\Delta\phi_0=n\pi$ ($n\in \mathbb{Z}$) and can be explained by the linearized model (blue line). The on-conservative force contributes to the total force for all other values of $\Delta\phi_0$, which leads to the amplification of the particle motion and a deviation of the normal mode splitting.}
\end{figure}

\begin{figure*}
	\centering
	\includegraphics[width=0.8\linewidth]{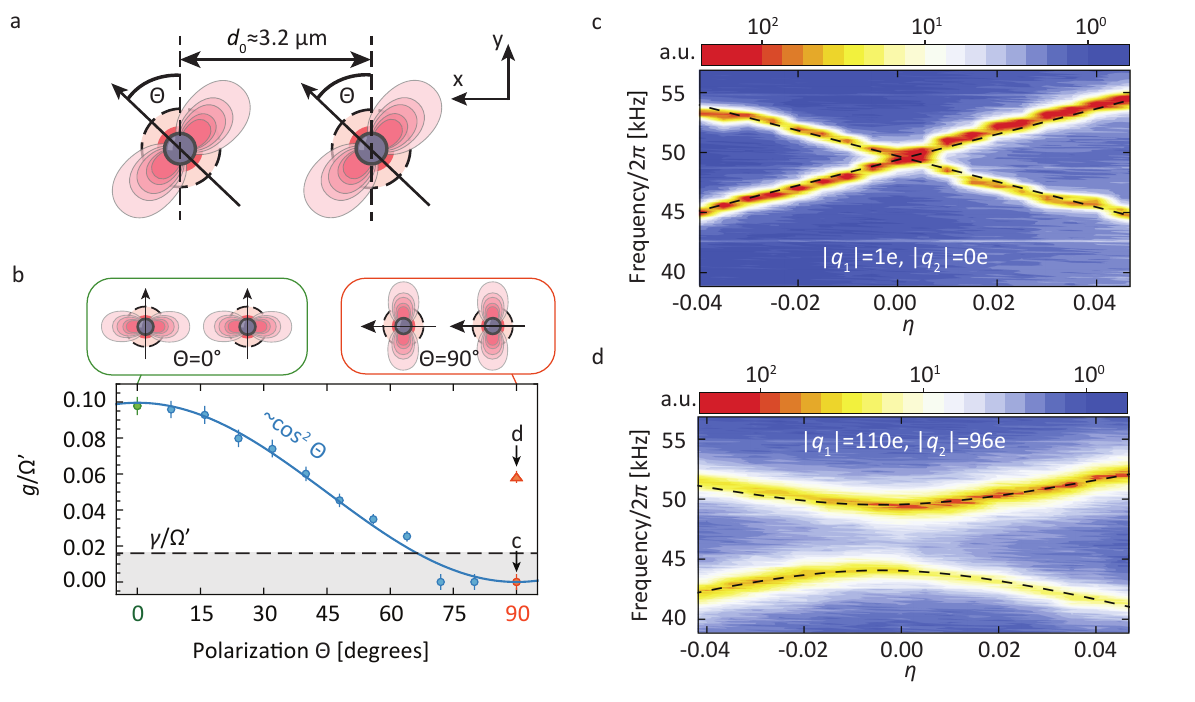}
	\caption{\label{fig:polarization} \textbf{Turning off dipole-dipole interaction to detect electrostatic interaction.} (a) For arbitrary polarization angle $\Theta$ the interference of electric fields is suppressed by $\cos^2 \Theta$. Two special cases of $\Theta=0^\circ$ and $\Theta=90^\circ$ are presented in green and orange rectangles, respectively. In the case of $\Theta=90^\circ$ there is no interference of the trapping and the scattered fields. (b) We measure the coupling rate due to the dipole-dipole interaction as a function of the polarization angle $\Theta$ between particles with an absolute number of $1\pm 1$ and $0\pm 1$ charges (circles). The interaction is maximal for the angle $\Theta=0^\circ$ (green circle). The avoided crossing is unresolved for coupling rates smaller than the mechanical linewidth $\gamma$ (gray region). At the angle $\Theta=90^\circ$ the dipole-dipole interaction is suppressed (orange circle). We observe coupling due to the electrostatic interaction between particles with $96\pm 21$ and $110\pm 24$ charges (orange triangle). (c) The avoided crossing is absent for horizontally polarized tweezers ($\Theta=90^\circ$) as the far-field dipole-dipole interaction is strongly suppressed. (d) An avoided crossing reappears for highly charged particles due to the strong electrostatic interaction.}
\end{figure*}

In order to obtain the coupling rate $g$, we measure the normal mode splitting as a function of $\eta$ (Fig. \ref{fig:nms}). In the experiment, we set the trap separation to $d_0\sim 3.15~\mu\text{m}$ and the optical phase difference to either $\Delta\phi_0=0$ (above) or $\Delta\phi_0=\pi$ (below) such that the interaction is purely conservative, but of either positive (attractive) or negative (repulsive) nature. The spectrogram exhibits an avoided crossing between the normal modes, which occurs for equal intrinsic mechanical frequencies ($\eta\equiv 0$) but only if $g>\gamma/2$. We conduct all measurements at pressures of around $1.5$ mbar, thus the avoided crossing is observable for coupling rates larger than $\gamma/2\pi\approx 1.5$ kHz. From here on we express the coupling rate in units of the modified mechanical frequency $\Omega'\approx \Omega+g$ as the ratio $g/\Omega'$ is independent of the optical power. We observe a frequency splitting of $\sim\pm 8$ kHz, which corresponds to a coupling rate of $g/\Omega'=\pm (0.09\pm 0.01)$. We can safely neglect several other coupling mechanisms in our investigations of the light-induced dipole-dipole interaction. We select particles with few charges, such that the additional coupling rate due to the electrostatic interaction was smaller than $|g_C|/\Omega'=(1.4\pm 0.6)\times 10^{-3}$ \cite{SI}. Hydrodynamic coupling is negligible as the ratio of the particle radius to the trap separation is small ($r/2d_0 < 0.05$) \cite{Svak2021Optica}.

We now demonstrate the full control over the conservative and non-conservative coupling rates. As the interaction arises from the interference between the trapping and scattered fields, we expect the coupling rate to oscillate with a period of $\lambda$ and decay with $d_0^{-1}$ due to the far-field nature of the dipole radiation at distances $d_0\gg \lambda$. To demonstrate this behavior, we measure the normal mode splitting for trap separations $d_0$ in the range of $ (2.2, 3.7) ~\mu\text{m}$ and for phase differences $\Delta\phi_0=0$ (blue points) or $\Delta\phi_0=\pi$ (orange points) in order to maximize the conservative interaction (Fig. \ref{fig:couplingscans}a). We observe a good fit of our theoretical model and the measured coupling rates in both cases \cite{SI}. At a distance of $d_0\approx 2.2~\mu\text{m}$ and for $\Delta\phi_0=0$ we observe the maximum coupling rate of $g/\Omega'=0.186\pm 0.017$. Effect of the dominant non-conservative interaction is apparent for the set trap separation of $d_0\approx 2.2~\mu\text{m}$ and the phase difference of $\Delta\phi_0=0.8 \pi$ (Fig. \ref{fig:couplingscans}b). The eigenfrequencies are degenerate for $\eta\in[0,0.07]$, from which we estimate the coupling rates of $g/\Omega\approx 0$ and $\bar{g}/\Omega\approx -0.068$. The constant pumping of energy into the system increases the particle motional amplitude by an order of magnitude. In order to demonstrate the dependence on the optical phase difference $\Delta\phi_0$, we measure the normal mode splitting at a fixed separation of $d_0\approx 3.2~\mu\text{m}$ (Fig. \ref{fig:couplingscans}c). We note that our linear model fails to fully predict the observed behavior. The actual interparticle distance is different from the trap separation due to the radiation pressure force of the dipole radiation. Moreover, in absence of an additional cooling mechanism, the particles are able to explore nonlinear terms in the interaction Hamiltonian, which affects the eigenfrequencies and modifies the normal mode splitting. Therefore, we observe a zero crossing due to the absence of the conservative forces at the phase of $\Delta\phi_0\approx 0.8\pi$, in agreement with our measurement in Fig. \ref{fig:couplingscans}b. In future work feedback cooling can be used to constrain the particle motion within linear dynamics.

Rotating the trapping laser polarization by an angle $\Theta$ from the y axis provides for another way to control the dipole-dipole interaction (Fig. \ref{fig:polarization}a). The magnitude of the dipole radiation along the x axis is smaller by a factor of $\cos \Theta$ due to the characteristic spatial profile of the dipole radiation in the far-field. The interference of the dipole radiation with the trapping field is suppressed by a factor of $\cos \Theta$ as a result of the scalar product of the two field components. Altogether this yields a decrease of the coupling rate by $\cos^2 \Theta$ in the far-field approximation, which is confirmed in the measurement in Fig. \ref{fig:polarization}b (circles and blue line). For the angle $\Theta=90^\circ$ the residual dipole-dipole interaction scales with $(kd_0)^{-3}$ due to the radial near-field component of the radiated field. We estimate the coupling rate of $g/\Omega'\approx 6\times 10^{-4}$ at $d_0\sim 3\lambda$, which we are unable to detect in the current experiment as $g/\gamma<10^{-2}$  (Fig. \ref{fig:polarization}c) \cite{SI}. Suppression of the dipole-dipole interaction allows us to explore electrostatic interaction between strongly charged particles. We trap particles with absolute charges $|q_1|/e=96\pm 21$ and $|q_2|/e=110\pm 24$ and opposite signs ($q_1q_2=-|q_1q_2|$) and observe a reappearance of an avoided crossing (Fig. \ref{fig:polarization}d). The measured coupling rate $g_C/\Omega'=0.058\pm 0.003$ fits well to the expected $g_C/\Omega'=0.047\pm 0.015$. Attractive interaction, reflected in $\Omega_-<\Omega_+$, is a confirmation that the charges are of opposite sign. Since the electrostatic coupling rate scales as $\propto d_0^{-3}$, in future experiments we will be able to resolve electrostatic interaction between particles with single charges at a distance of $d_0\sim 2 ~\mu \text{m}$ and at pressures below $10^{-3}$ mbar. Altogether our platform allows for exploring hybrid schemes with both dipole-dipole and electrostatic interactions. 

In conclusion, we demonstrate controllable (attractive and repulsive) light-induced dipole-dipole interaction between two silica nanoparticles levitated in distinct optical traps with coupling rates up to $20\%$ of the mechanical frequency. We thereby significantly expand the toolbox of optical binding by trapping in a phase-coherent optical tweezers array, which will enable further studies of optical interactions between Rayleigh particles \cite{ZemanekRMP} or atoms \cite{Ritschtwoatoms,Kaiseratoms,Hasslinger} at sub-wavelength distances \cite{ForbesNanophotonics}. Furthermore, we are able to tune conservative and non-conservative interactions, which allows for non-reciprocal interactions between the particles. The optical interaction can be effectively switched off, which allows us to tune in electrostatic interaction between two charged particles. In stark contrast to cavity-mediated entanglement schemes between two mechanical oscillators \cite{MercierEntanglement,Kotler2021,SticklerEntanglement,CernotikEntanglement,Guerreiro}, here the demonstrated direct interaction mechanisms will allow for stationary entanglement between particle motions in the future. The creation of entanglement will depend only on the coupling rate $g/\Omega'$ and the average occupation of the CoM normal mode $\langle n_+\rangle$ \cite{HartmannPlenio,LudwigEntanglement,QvarfortEntanglement}, which includes coupling to the environment and thus could be used as a probe of the decoherence in the quantum-to-classical transition \cite{BenedettiProbe}. Suppression of the dipole-dipole interaction and discharging of particles may allow for investigation of weak forces in the quantum regime, such as the Casimir-Polder force \cite{WeissDelocalization,PlenioDrivenEntanglement,CasimirPhonon2019}. Engineered strong optical and electrostatic interactions between multiple macroscopic objects may open up many research avenues in quantum physics. We foresee that the platform described in this work -- with a possible addition of an optical cavity -- can be used for quantum simulation with mechanical degrees of freedom \cite{LudwigMarquardt,HeinrichMarquardt,Holzmanncoupled2021}, enhanced quantum sensing \cite{ClerkSensing}, collective effects \cite{Xuereb2012,DomokosRitsch,RitschBroadband}, (quantum) synchronization \cite{Sadeghpour,Fazio,Bruder}, studies of molecular structures \cite{Dursomolecule}, ultrastrong coupling between harmonic oscillators \cite{MarkovicUltrastrong} or phonon transport and thermalization \cite{TongcangParticleChain}.

\begin{acknowledgments}
	\textit{Acknowledgments. } We thank Oto Brzobohat\' y, Pavel Zem\' anek, Helmut Ritsch and Oriol Romero-Isart for insightful discussions. UD would like to thank Vladan Vuleti\' c for initial discussions about trap arrays. This research was funded in whole or in part by the Austrian Science Fund (FWF, Project No. I 5111-N and START Project TheLO, Y 952-N36), the European Research Council (ERC 6 CoG QLev4G), by the ERA-NET programme QuantERA under the Grants QuaSeRT and TheBlinQC (via the EC, the Austrian ministries BMDW and BMBWF and research promotion agency FFG), by the European Union’s Horizon 2020 research and innovation programme under Grant No. 863132 (iQLev).  MAC acknowledges support from the FWF Lise Meitner Fellowhip (M2915, "Entropy generation in nonlinear levitated optomechanics"). HR, KH, BAS acknowledge funding from the Deutsche Forschungsgemeinschaft (DFG, German Research Foundation) under Grant No. 439339706. For the purpose of Open Access, the author has applied a CC BY public copyright license to any Author Accepted Manuscript (AAM) version arising from this submission. 
	
\end{acknowledgments}

\onecolumngrid
\appendix
\newpage
\makeatletter
\renewcommand*{\@biblabel}[1]{\hfill#1.}
\makeatother

\setcounter{figure}{0}
\setcounter{equation}{0}
\renewcommand{\thefigure}{S\arabic{figure}}
\renewcommand{\theequation}{S\arabic{equation}} 

\section{\Large Supplementary material}

\subsection{Theory basics}

At this point we will sketch the derivation of the equations of motion and the coupling constants. The force on a dipole is given by the gradient of the scalar product of its dipole moment with the incident electric field, where the gradient only acts on the electric field, treating the dipole moment as a constant. We can express the dipole moments inside of a $j$-th dielectric particle via the particles polarizability $\alpha_j=\varepsilon_0 V_j \chi$ and the drive (trapping) field $\mathbf{E}_0(\mathbf{r})$, with the particle volume $V_j$ and $\chi=3(\varepsilon-1)/(\varepsilon+2)$ with the permittivity $\varepsilon$. The incident field at the position $\mathbf{r}_j$ away from a particle is a sum of $\mathbf{E}_0(\mathbf{r})$ and the scattering fields of all other particles, which can be expressed by the Green's tensor $\mathbf{G}(\mathbf{r})$ of the transverse Helmholtz equation:

\begin{align}
	\mathbf{G}(\mathbf{r}) = \frac{e^{ikr}}{4\pi}\left[ \frac{3\mathbf{r}\otimes\mathbf{r} - r^2}{r^5}(1 - ikr) + k^2 \frac{r^2 - \mathbf{r}\otimes\mathbf{r}}{r^3} \right],
\end{align}
with $r=|\mathbf{r}|$ and where we omitted the unity matrix for simplicity.

By expanding the force to the second order in the particle volume, one can identify three contributions to the total force acting on one particle. The first is a sum of the well-known gradient force and the non-conservative radiation pressure originating only from the trapping field. The second contribution comes from the scattering fields of the other particles acting on the dipole moment due to the laser field, while the third contribution can be interpreted as the laser acting on the dipole moment induced by the scattering fields of the other particles. The second and third contribution together condense in what is called the optical binding force:

\begin{align}
	\mathbf{F}_j^{\rm bind} = \nabla_j \text{Re} \left[ \sum_{j'\neq j} \frac{\alpha_j \alpha_{j'}}{2\varepsilon_0} \mathbf{E}^*_0(\mathbf{r}_j)\cdot \mathbf{G}(\mathbf{r}_j - \mathbf{r}_{j'})\mathbf{E}_0(\mathbf{r}_{j'}) \right],
\end{align}
with the nabla operator $\nabla_j$  acting on the coordinate $\mathbf{r}_j$. Expanding the forces to the third order in the particle volume would add the gradient force of scattering fields alone, as well as the interaction of the trapping field with the second order scattering fields. However, we will neglect these higher order terms ($\sim \mathcal{O}(V^3)$) in this work.

The coupling rates used in this work can be derived by expanding the optical binding forces on the particles around their respective equilibrium positions given by the gradient force of the trapping field. Here, we approximate the trapping fields as focused Gaussian beams. We take only the contributions to the optical force that are $\propto d_0^{-1}$. We consider two spherical dielectric particles of equal mass, each trapped in an optical trap with frequencies $\Omega_j$ along the optical axis and optical phases $\phi_j$. 

The linearized dynamics along the optical axis with particle coordinates $z_{1,2}$ is then, in leading order of the trap separation $d_0$, determined by the coupling constants $k_1$ and $k_2$,
\begin{eqnarray}
		m\ddot{z}_1 &=&- \left( m\Omega_1^2 + k_1 + k_2 \right) z_1 + (k_1 +k_2)z_2 \nonumber\\
		m\ddot{z}_2 &=& - \left( m\Omega_2^2 + k_1 - k_2 \right) z_2 + (k_1-k_2)z_1.\label{lang}
\end{eqnarray}

The coupling constant $k_1$ describes the conservative part of the optical binding forces, while $k_2$ describes a non-conservative interaction, as indicated by the opposite sign in the equations of motion. They depend on the relative local phase $\Delta\phi_0 = \phi_1-\phi_2$ and the tweezer separation $d_0$ as $k_1 = G \cos(kd_0) \cos(\Delta\phi_0)/kd_0$ and $k_2 = G \sin(kd_0) \sin(\Delta\phi_0)/kd_0$. Constant $G= \alpha^2 k^5 \sqrt{P_1 P_2}/(2c w_0^2\pi^2\varepsilon_0^2)/m$ is a positive function of the particle polarizability $\alpha$ and optical powers $P_{1,2}$, where $w_0$ is the trap waist, $\varepsilon_0$ is the vacuum permittivity and $c$ is the speed of light. The distance $d_0$ and the relative tweezer phase $\Delta\phi_0$ allow tuning between purely conservative and non-conservative interactions.

The non-conservative contribution to optical binding emerges from the radiation pressure induced by the scattered fields, which constantly pumps energy into the system, thus it can't be derived from a Hamiltonian. This is also observed with curl-forces \cite{berry2013}, in ro-translational oscillators \cite{AritaBinding2,Nanowires} or for binding of particles of different sizes \cite{KarasekAssymetric,DogariuAssymetric,ChvatalAssymetric}. The equations of motion along the x and y axes follow the same form.

\subsubsection{Eigenfrequencies of the coupled system}

We diagonalize Eqs. \ref{lang} in order to obtain the eigenfrequencies of the normal modes for arbitrary intrinsic mechanical frequencies $\Omega_1$ and $\Omega_2$:
\begin{equation}
\Omega_{\pm}^2=\frac{1}{2}\left(\Omega_2^2+\Omega_1^2+2k_1/m\mp\sqrt{(\Omega_2^2-\Omega_1^2)^2+4(k_1/m)^2-4(\Omega_2^2-\Omega_1^2)k_2/m}\right).
\end{equation}
The splitting is minimal for $\Omega_2^2-\Omega_1^2=2k_2/m$. This is reached for the control parameter $\eta_{\rm m}=k_2/(m\Omega^2)$, where $\Omega$ is the mean mechanical frequency for $\eta=0$. For small $k_1\text{, }k_2\ll m\Omega_1^2$ the eigenfrequencies are:
\begin{equation}
\Omega_\pm=\sqrt{\Omega_1^2+(k_1+k_2)/m\mp\sqrt{k_1^2-k_2^2}/m}\approx \Omega_1+\frac{(k_1+k_2)/m\mp\sqrt{k_1^2-k_2^2}/m}{2\Omega_1},
\end{equation}
with the minimal splitting of:
\begin{equation}
\Omega_--\Omega_+\approx \frac{\sqrt{k_1^2-k_2^2}}{m\Omega\sqrt{1-\eta_{\rm m}}}.
\end{equation}
The conservative interaction has to be larger than the non-conservative interaction $k_1^2>k_2^2$ in order to have an avoided crossing. In the case of purely conservative interaction ($k_2\equiv 0$) the splitting is minimal for $\Omega_2=\Omega_1\equiv \Omega$:
\begin{eqnarray}
\Omega_{\pm}=\sqrt{\Omega^2+(k_1\mp k_1)/m} &\Rightarrow& \Omega_+=\Omega\text{, }\Omega_-\approx \Omega+\frac{k_1}{m\Omega}\nonumber\\
\Omega_--\Omega_+&\approx& \frac{k_1}{m\Omega}=2g.
\end{eqnarray}

\subsubsection{Interaction suppression by polarization}

The radiated electric field for an arbitrary polarization angle $\Theta$ consists of a radial and an azimuthal contribution:
\begin{eqnarray}
E_R(R)&=&-E_0\frac{\alpha k^2\sin\Theta}{4\pi\varepsilon_0 R}e^{ikR}\left(\frac{2}{k^2R^2}-\frac{2i}{kR}\right)\nonumber\\
E_\varphi(R)&=&-E_0\frac{\alpha k^2\cos\Theta}{4\pi\varepsilon_0 R}e^{ikR}\left(\frac{1}{k^2R^2}-\frac{i}{kR}-1\right).
\end{eqnarray}
If the light is polarized along the x axis ($\Theta=90^\circ$), the azimuthal component of the radiated field disappears $E_{\varphi}\equiv 0$. However, the near-field radial component $E_R$ yields the following conservative coupling rate:
\begin{equation}
g_{\text{near}}(d_0,\Delta\phi_0=0)=\frac{G}{2\Omega}\left(-\frac{2}{k^3d_0^3}\cos(kd_0)+\frac{2}{k^2d_0^2}\sin(kd_0)\right).
\end{equation}
For $d_0\sim 3\lambda$ the coupling rate is a factor of $2/(36\pi^2) \approx 5.6\times 10^{-3}$ of the coupling rate for $\Theta=0^\circ$.

\subsubsection{Electrostatic interaction}

Electrostatic interaction between dielectric objects is purely conservative with the interaction energy:
\begin{equation}
H_{C}=\frac{1}{4\pi\varepsilon_0}\frac{q_1q_2}{\sqrt{(d_0+x_1-x_2)^2+(y_1-y_2)^2+(z_1-z_2)^2}},
\end{equation}
where $q_1$ and $q_2$ are particle charges, $d_0$ is the trap separation and particle motions $x_{1,2}$, $y_{1,2}$ and $z_{1,2}$. We expand the Hamiltonian to the second order in $z_{1,2}$ and obtain:
\begin{equation}
H_{C}=\frac{1}{4\pi\varepsilon_0}\frac{q_1q_2}{d_0}\left(1-\frac{(z_1-z_2)^2}{2d_0^2}\right)\equiv \frac{k_1}{2}(z_1^2+z_2^2)-k_1z_1z_2,
\end{equation}
where we have defined $k_1=-\frac{q_1 q_2}{4\pi\varepsilon_0 m}\frac{1}{d_0^3}$ in analogy to the conservative optical interaction. The coupling rate due to the electrostatic interaction is given by:
\begin{equation}
g_C=\frac{k_1}{2m\Omega'}=-\frac{q_1 q_2}{8\pi\varepsilon_0 m\Omega'}\frac{1}{d_0^3},
\end{equation}
where $\Omega'=\sqrt{\Omega^2-\frac{q_1 q_2}{8\pi\varepsilon_0 m d_0^3}}$ is the modified mechanical frequency due to the electrostatic interaction.

\subsection{Experimental setup}
\begin{figure}[!ht] 
	
	\includegraphics[width=\linewidth]{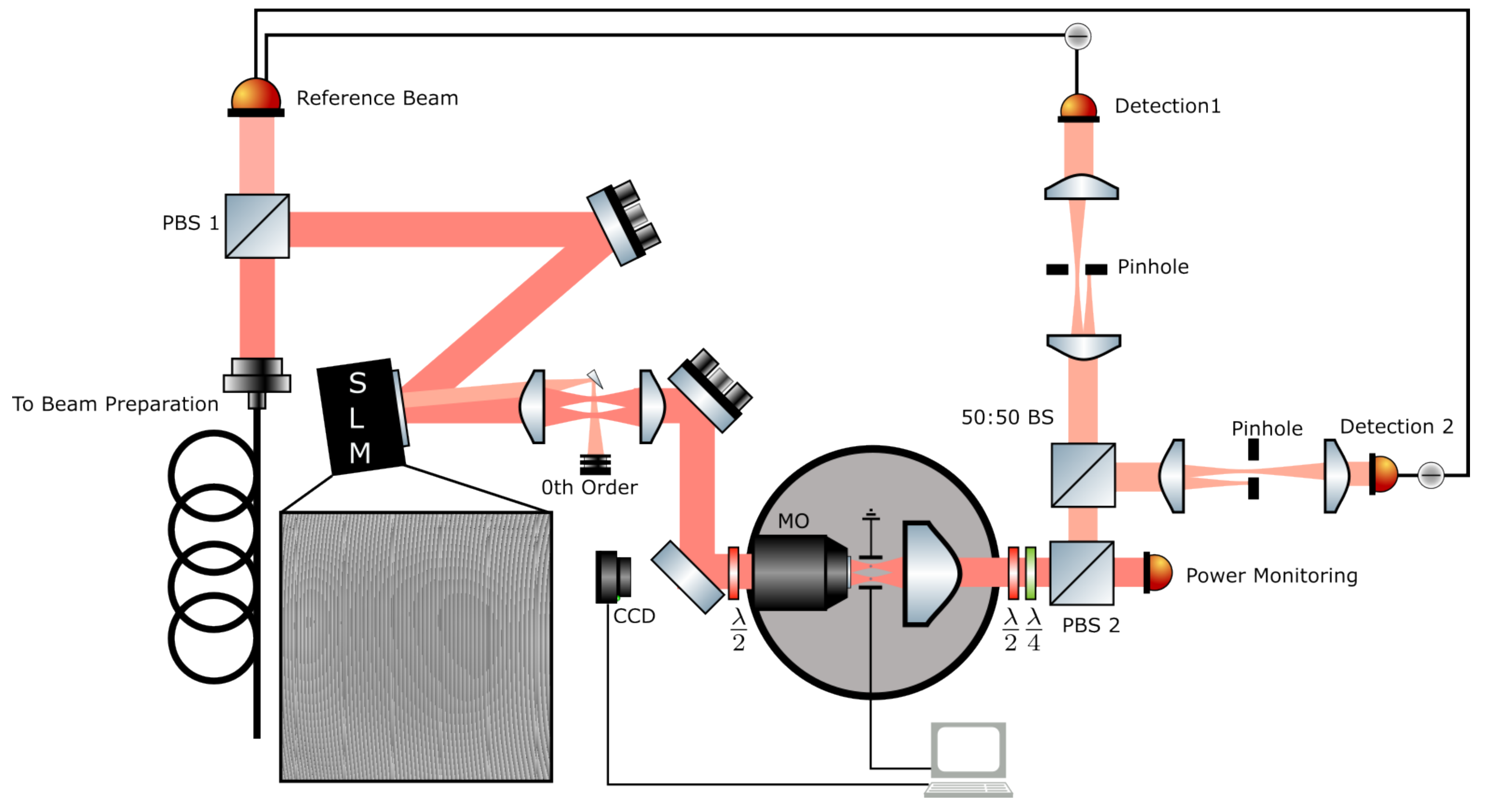}
	\caption{Setup for generating multiple traps using a spatial light modulator. 1064 nm laser light, with a diameter of 8.7mm, is reflected on a Spatial Light Modulator (SLM Meadowlark Inc.) where a phase profile is imprinted. In the Fourier plane of the trapping objective the phase profile is translated into an amplitude profile with correct imaging being ensured via a 4f configuration using a 1:1 telescope. The light is recollimated and split for detection, with the traps being selected using movable pin holes in 1:1 telescopes.}
	\label{Schematic}
\end{figure}

The experimental setup is shown in Fig. \ref{Schematic}. The trapping beam ($\lambda=1064$ nm,  Keopsys fiber amplifier seeded by a Mephisto laser) is expanded to a diameter of $8.7$ mm in order to overfill the apertures of the spatial light modulator (SLM, Meadowlark Inc.) and the microscope objective. We imprint a phase profile with the SLM into the trapping beam that transforms into an amplitude profile in the Fourier plane of the trapping optics. The phase profile is calibrated to compensate for aberrations, nonlinearities of the SLM response and non-flatness of the SLM surface. Inset of Fig. \ref{Schematic} shows the phase profile used to generate two traps spaced by $3.4~\mu\text{m}$. This phase profile is imaged onto the trapping optics using a 1:1 telescope set in a 4f configuration (focal length of lenses $300$ mm). The trap is then generated using a microscope objective (NA$=0.8$, WD $= 1$mm, Nikon Corp.), focusing the beam to two traps of waist $\approx 730nm$. The total power used in front of the vacuum chamber is $\approx 1.2$ W. We maintain a stable pressure of $p\approx 1.5$ mbar in the vacuum chamber, at which a single particle is in a thermal equilibrium with the environment.

A pair of electrodes mounted on a 3D piezo stage (MX25, Mechonics Inc.) is placed around the beam focus, with a spacing of $D_c=(230\pm 15)~ \mu\text{m}$. The light is recollimated using an aspheric lens (C660TME-C, Thorlabs) and split at a polarization beamsplitter (PBS2) for power monitoring. The reflected arm is split with a $50:50$ beamsplitter. We mount a pinhole on a translation stage in the focus of a 1:1 telescope ($f=125$ mm) in each of the beamsplitter outputs, which we use to select individual laser beams in order to separate the detection. In each arm the trapping beam goes onto one photodiode of a balanced photodetector (PDB425C, Thorlabs), while the reference beam of equal power (taken from PBS1) is focused onto the other photodiode in order to suppress the intensity noise. We acquire $2$ seconds of data at a sampling rate of $2.5~\text{MSa}/\text{s}$ with an oscilloscope (PicoScope 5444D).

\subsection{Charge calibration}

In order to estimate the magnitude of the electrostatic interaction between the particles, we performed charge calibration for each particle used in the experimental runs. We increase the trap separation to approximately $18~\mu\text{m}$ in order to minimize all interactions and cross-talk in the detection. We subsequently apply a sinusoidal voltage ($F_d(t)=F_0 \sin (\Omega_d t)$) to the electrodes set along the x axis, with the driving frequency $\Omega_d$ close to the mechanical frequencies along the x axis $\Omega_x$. We measure the particle displacement as a function of the driving frequency \cite{Magrini2}:
\begin{equation}
	\label{1}
	\langle x_d^2\rangle =\frac{\langle F_d^2\rangle}{m^2(\Omega_{x}^2-\Omega_{d}^2)^2}
\end{equation}
Here, $\langle \cdot \rangle$ is the time average, $\langle F_d^2\rangle= F_0^2/2$ is the applied half-amplitude force, $m=(7.0\pm 0.7)$ fg is the particle mass, $\Omega_x$ is the trap frequency (in rad/s) along the x axis, $\Omega_d$ is the drive frequency (in rad/s).   

Assuming a model for a massive point charge in a parallel-plate capacitor, we get $F_0 = \frac{qV}{D_c}$ where $q=Ne$ is the particle charge, $V$ is the applied voltage and $D_c=(230\pm 15)~\mu\text{m}$ is the distance between the electrodes. We can express Eq. \ref{1} in terms of the number of charges to get:
\begin{equation}
	N = \frac{\sqrt{2}mD_c}{eV}\left|\Omega_{x}^2-\Omega_{d}^2\right|\sqrt{\langle x_d^2\rangle}
\end{equation}
After a position-displacement calibration \cite{NovotnySensing2018}, we extract $\langle x_d^2\rangle$ by integrating over the drive frequency in the spectrum. Table \ref{table1} lists the number of charges for each particle used in the main text. 

\begin{figure}[!ht] 
	\label{2}
	\includegraphics[width=\linewidth]{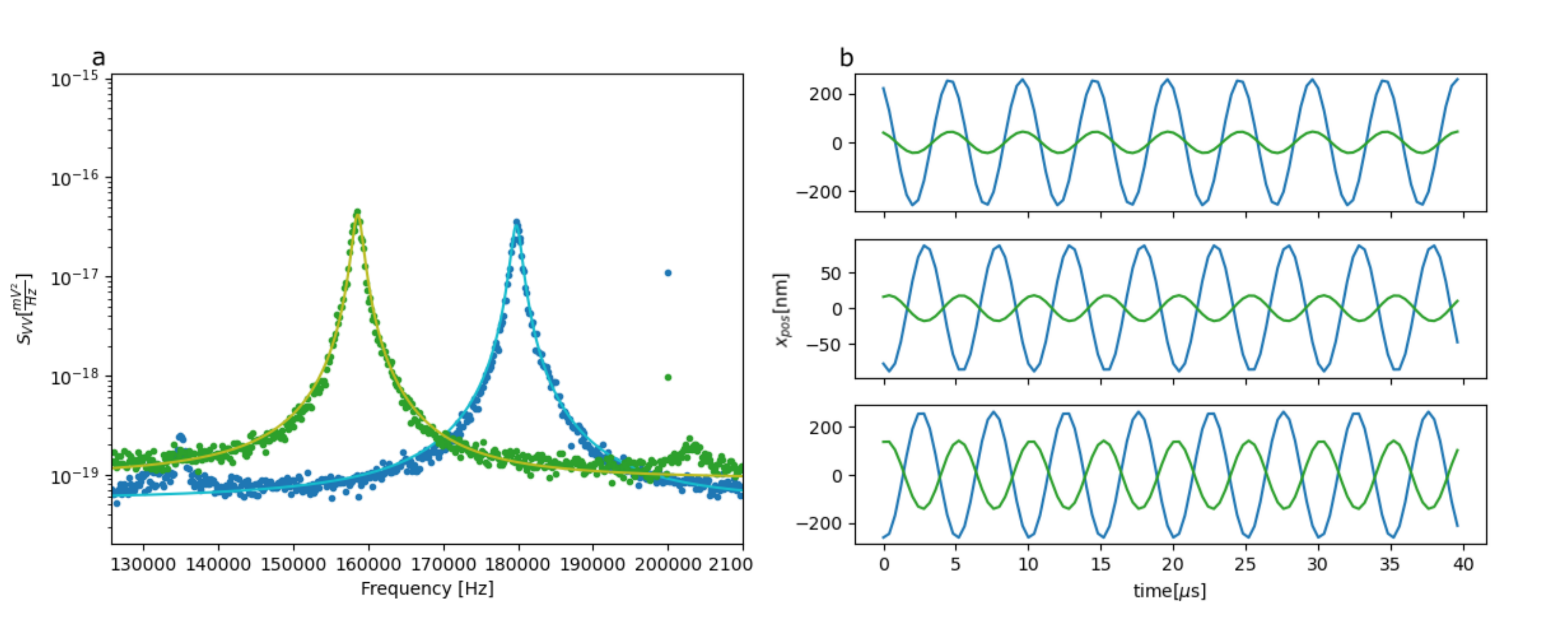} 
	\caption{a) Example spectra with the drive frequency $\Omega_d=2\pi\times200$ kHz. The mechanical frequencies are set to be different in order to check that we have independent readout of the particles. b) Time trace of the driven particle motion at the drive frequency. The phase between the particles is indicative of the sign of the charges. Equal and opposite sign of charges is reflected in the in-phase or out-of-phase response to the drive, respectively. }
	
\end{figure}

\begin{table}[h!]
	\begin{tabular}{ |c|c|c|c|c| } 
		\hline
		Figure in the main text & $|N_1|$ & $|N_2|$& sign($N_1 N_2$)&$g_C/\Omega'$\\ 
		\hline
		3a & $23\pm5$ &$5\pm 2$ &1&$-(4.6\pm 2.1)\times 10^{-4}$\\ 
		\hline
		3b & $3\pm 1$ & $1\pm 1$ &-1&$(1.2\pm1.3)\times 10^{-5}$\\ 
		\hline
		4c & $1\pm 1$ & $0 \pm 1 $&n.a.&$(0\pm4)\times 10^{-6}$\\ 
		\hline
		4d & $110 \pm 24$ & $96 \pm 21$ &-1&$0.047 \pm 0.015$\\ 
		\hline
	\end{tabular}
	\caption{Results of the charge calibrations performed on 4 sets of particles. $N_1$ and $N_2$ are the number of elementary charges on the particles. Each row represents a set of particles used for measurements in the main text. We provide the coupling rate due to the electrostatic interaction at a distance of $3.2~\mu\text{m}\approx 3\lambda$ and for the mechanical frequency $\Omega=2\pi\times 50$ kHz.}
	\label{table1}
\end{table}

\subsection{Interaction model used for fitting}

We correct for aberrations \textit{in situ} with the SLM, however the trapping field still has the shape of an Airy function in the focus due to the high numerical aperture of the microscope objective. This leads to a small overlap of the trapping fields at distances larger than the trap waist, therefore we are unable to separate the trapping fields in the description of the total interaction. This leads to a "self-interference" effect, where for example at the position of particle 2 the scattered field of particle 1 interferes with the tail of the trapping field for particle 1 \cite{WeiLateralBinding}. We model this with a standing wave in the focal plane with a relative electric field magnitude $A$. Already a weak tail of the trapping potential can have a large impact as it becomes comparable to the magnitude of the dipole radiation. This effect leads to a slight modification of the trap positions.

Furthermore, the radiation pressure of the scattered fields displaces the particles from the desired trap positions. The actual distance between the particles is smaller (larger) in the presence of an attractive (repulsive) force. This is confirmed when we compare coupling rates obtained for $\Delta\phi_0=0$ and $\Delta\phi_0=\pi$ in Fig. 3a in the main text; the period between the zero crossings of the coupling rate is larger for positive coupling rates, which is a result of the displacement by the optical force.

We include both effects in the model of the dipole-dipole interaction and find an excellent fit between the model and the experimental data for distances larger than $\sim 2.4~\mu\text{m}$ in Figure 3a in the main text. We point out that at smaller trap separations we neglect several features of the trap potential and the near-field optical interaction, as well as the (small) effect of the aerodynamic coupling. In future experiments at closer trap separations we will have to investigate the trap potential shape, as well as include the dipole radiation component $\propto d_0^{-2}$ which is non-negligible at trap separations $d_0\sim \lambda$.

\subsection{Normal mode splitting of the x and y motions}

\begin{figure}[!ht] 
	\includegraphics[width=0.4\linewidth]{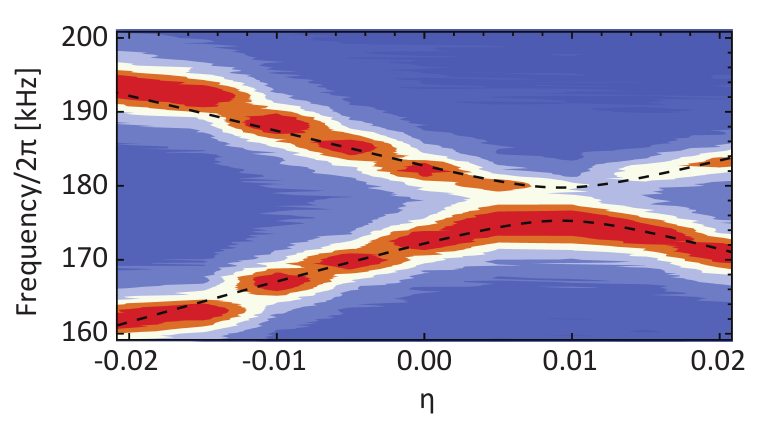} 
	\caption{Normal mode splitting of the x motion due to the dipole-dipole interaction between two particles at a distance $d_0\approx 2.2~\mu\text{m}$.}
	\label{nmsx}
\end{figure}

In the main text we have presented the results obtained only for the motion along the optical axes (z axes). However, standard optical binding interaction exists for all three directions of the particle motion due to the modification of the interparticle distance $d^2=(d_0+x_1-x_2)^2+(y_1-y_2)^2+(z_1-z_2)^2$. The coupling rate between the x motions has the following form: $g_x\sim G/(2\Omega_x kd_0))$. We observe an avoided crossing between the x motions at a distance $d_0\approx 2.2~\mu\text{m}$ (Fig. \ref{nmsx}). We extract the coupling rate of $g_x/\Omega'_x=0.013\pm 0.002$, which is significantly smaller than the coupling rate between the z motions from the main text as expected. The coupling rate $g_x/\Omega'_x$ scales $\propto \Omega_x^{-2}$, thus the higher mechanical frequency by a factor of $\Omega_x/\Omega\approx 4$ yields a smaller ratio $g_x/\Omega_x$ by a factor of $\sim 16$ in comparison to $g/\Omega$, which fits to the measured value of $g/\Omega'\approx 0.186$ from the main text. We are unable to observe the avoided crossing between the y motions as the coupling rate is smaller than the mechanical damping.

\end{document}